\documentclass[aps,prc,twocolumn,showpacs,superscriptaddress,groupedaddress]{revtex4}

\usepackage{graphicx}
\usepackage{color}

\begin{document}

\preprint{1}

\title{Extraction of dihadron-jet correlations with rigorous flow-background subtraction
in a multiphase transport model}

\author{Yuhui Zhu}
\affiliation{Shanghai Institute of Applied Physics, Chinese Academy of Sciences, Shanghai 201800, China}
\affiliation{University of Chinese Academy of Sciences, Beijing 100049, China}

\author{Y. G. Ma\footnote{Author to whom all correspondence should be addressed:
ygma@sinap.ac.cn}}
\author{J.  H. Chen}
\author{G. L. Ma}
\author{S. Zhang}
\author{C. Zhong}



\affiliation{Shanghai Institute of Applied Physics, Chinese Academy of Sciences, Shanghai 201800, China}


\date{\today}

\begin{abstract}
Dihadron azimuthal correlations in Au+Au collisions at $\sqrt{S_{NN}}$=200 GeV have been
explored by using a multi-phase transport (AMPT) model. In order to obtain the contributions
from jet-medium interactions,
the combined harmonic flow background is subtracted from the raw
dihadron correlation functions. The signals  are compared in three associated
transverse momentum ($p_{T}^{assoc}$) bins: 0.2-0.8 GeV/c, 0.8-1.4 GeV/c and
1.4-2.0 GeV/c from central to semi-peripheral geometries. The medium modifications
are observed from changes in the signal shape and the relative jet
contribution has been obtained within the change in the centrality from peripheral to central one.
A strong $p_{T}^{assoc}$ dependence of the RMS width of jet correlation function is
observed within the central geometry bin, i.e. 0-10$\%$.

\end{abstract}

 \maketitle

 \section{Introduction}
 Lattice quantum chromodynamics (QCD) calculations predicted a phase transition from hadron
 gas to a deconfined matter in ultra-relativistic heavy ion
collisions~\cite{QCD-phasetran,White-papers}.  A hot and
dense partonic matter, called Quark-Gluon
Plasma (QGP), is formed in these collisions and is found to be strongly interacting.

In the literature, jet quenching~\cite{jet-quenching} has been used as a signal of the QGP phase
formation since it is interpreted as the evidence for
the interaction between jets and the QGP medium. Dihadron azimuthal correlation
acts as  a probe for the study of jet quenching in high-energy heavy ion collisions.
During the study of dihadron azimuthal correlation functions,  an away-side double peak structure is observed in RHIC
experiments at relatively high transverse momentum ($p_T$) \cite{soft-soft-ex,sideward-peak2,dihadron-2009}, which is
regarded as an evidence for jet quenching.

There are ongoing efforts within the RHIC comunity to study the inner mechanism of the jet-medium interaction through
the dihadron correlation function. In this contest, many theoretical models have been put forward time to time
such as shock wave model~\cite{Stoecker}, gluon radiation model~\cite{Koch},
medium-induced gluon bremsstrahlung~\cite{large-angle,opaque-media-radiation}, waking
the colored plasma and sonic Mach cones~\cite{Ruppert}, sonic
booms and diffusion wakes in thermal gauge-string
duality~\cite{sonic-booms}, jet deflection \cite{deflection} and
strong parton cascade mechanism etc~\cite{ma-sqm,di-hadron,three-hadron,time-evolution,pt-dependence}.

From the above studies, we  know that dihadron azimuthal correlation signal
is the key point to a solid physical interpretation. However, constructing a precise dihadron correlation background is
a complex task.  This is mainly because of the contributions from the odd-order harmonic
flows, which are induced by initial geometry fluctuations, to the away-side double peak
structure \cite{Ko,GL}. Theoretically, one can extract the harmonic
flow coefficients and then reconstruct the high order harmonic background.
         However, due to the  overestimation or underestimation of the non-flow
         contributions in our calculation method, it is hard to
         get an ideal flow coefficient. Therefore, more detailed analysis concerning the background construction
         is the need of the present time.

In this paper, we will employ two different methods for calculating background, which are discussed
in detail in the upcoming section. These methods will be applied for the study of transverse momentum ($p_{T}$) and centrality dependences
(0-10\%, 20-40\% and
50-80\%) of dihadron azimuthal correlation functions in 200 GeV/c Au+Au collisions.
Finally, the emphasis will be put on the study of  jet-induced signals in different associated $p_T$
($p_{T}^{assoc}$) bins from the central collisions.

The paper is organized as follows. Section II gives a
brief introduction of the simulation model. Section III describes the two
different analysis methods, which are used for background construction in the
study of  dihadron azimuthal correlation functions. The
results and discussions are given in  Section  IV, which is followed by
a summary in Section V.

\section{Methodology: A Multi Phase Transport Model (AMPT)}

A multi-phase transport model
(AMPT)~\cite{AMPT} is employed for the study of dihadron azimuthal
correlations. The model has four main components for describing the
physical processes in relativistic heavy-ion collisions: 1) the initial
conditions from HIJING model~\cite{HIJING}, 2) partonic
interactions modeled by Zhang's Parton Cascade model (ZPC)~\cite{ZPC},
3) hadronization, and 4) hadronic rescattering simulated by A
Relativistic Transport (ART) model~\cite{ART}.
The model works as follows: first, many excited strings initiated from HIJING are melted into
partons in the string melting version of AMPT model ~\cite{SAMPT}(abbr. ``the Melting AMPT version").
Then, a simple quark coalescence model is used to combine the
partons into hadrons. On the other hand, in the default version of AMPT
model~\cite{DAMPT}(abbr. ``the Default AMPT version"),
minijet partons are recombined with their parent strings, when they
stop interactions and the resulting strings are converted to
hadrons via the Lund string fragmentation model~\cite{Lund}. It indicates that the
Melting AMPT version undergoes a partonic phase much more than
 the Default AMPT version. The details of the AMPT
model are available in a review paper~\cite{AMPT} and previous
works~\cite{AMPT,SAMPT,Jinhui}.

For the present study,  the Melting AMPT version is used to perform the
simulation for 200 GeV/c Au+Au collisions. Moreover, in order to concentrate on partonic stage interactions, final
hadronic rescattering process is turned off as well.

\section{Analysis Method}

The raw dihadron azimuthal correlations are obtained by following the
experimental procedure~\cite{soft-soft-ex,sideward-peak2}.  In the experiments,  the
azimuthal correlation is established between a high $p_{T}$ particle (trigger
particle) and low $p_{T}$ particles (associated particles). In the present work,
the particles in the range $p_T >$ 2.5 ($\le 2.5$) GeV/c  are specified as  trigger particles (associated particles).
Both the trigger and associated particles are required to be within a pseudo-rapidity
window of $|\eta|<1.0$. The raw signal is obtained by accumulating
pairs of trigger and associated particles into $\Delta\phi =
\phi_{assoc} - \phi_{trig}$ distributions in the same event.

\begin{figure*}[hbtp]
\includegraphics[width=5.2cm]{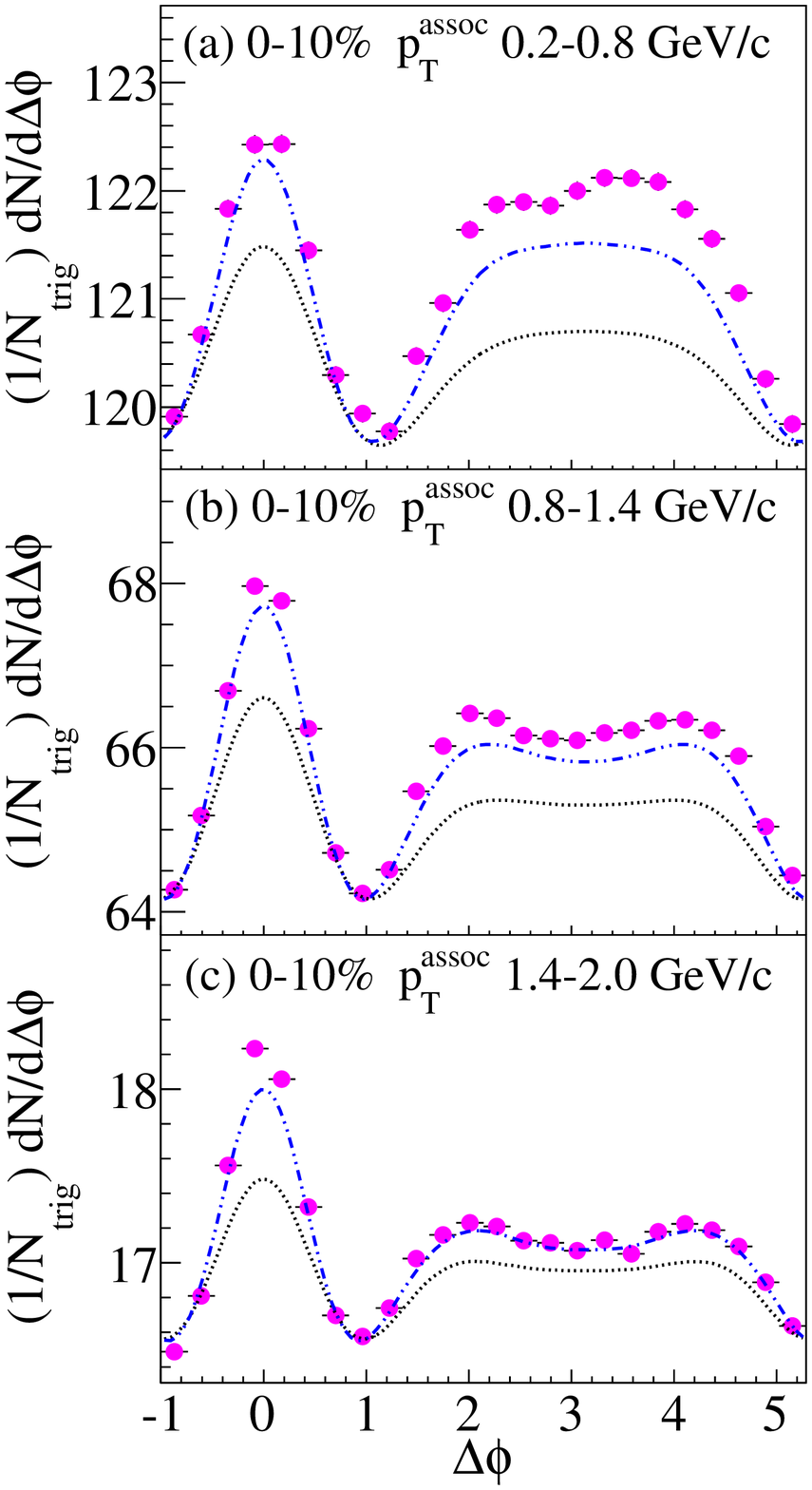}
\includegraphics[width=5.2cm]{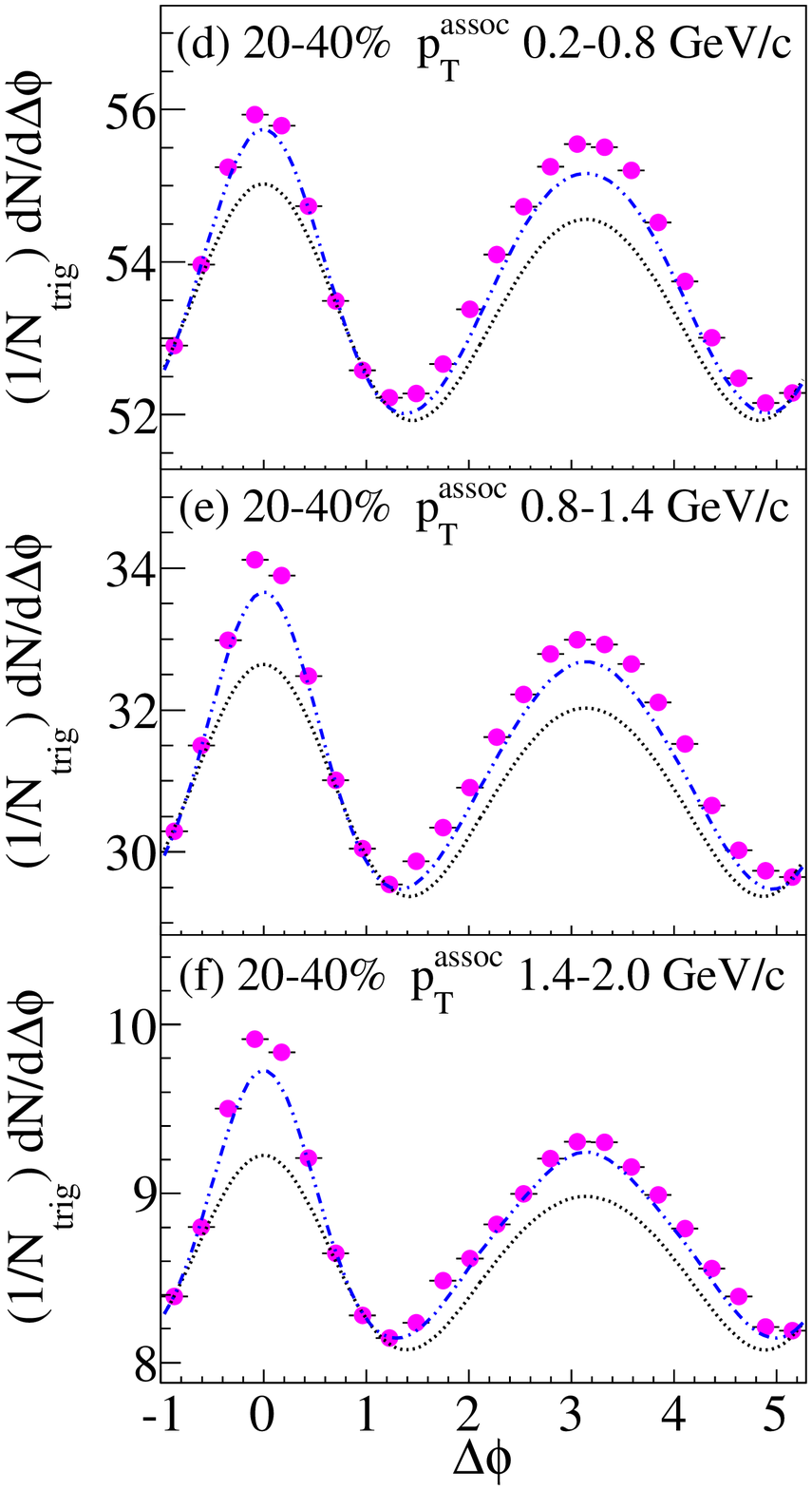}
\includegraphics[width=5.2cm]{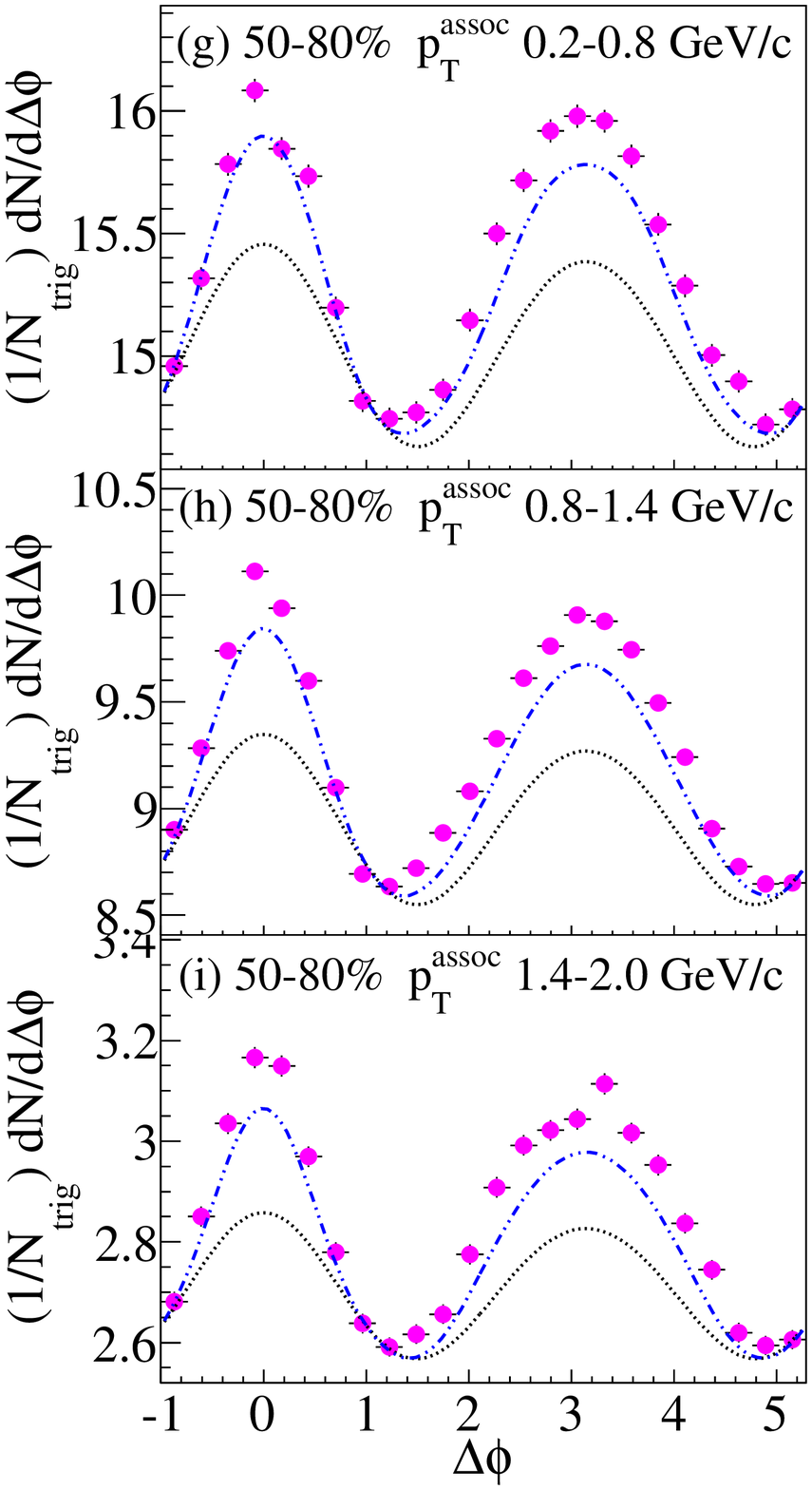}
\caption{(Color online) Dihadron correlation functions in 200 GeV/c
Au+Au collisions derived in three $p_{T}^{assoc}$ bins as illustrated in the
    plots.
 (Left)  0-10\% centrality,
 (Middle)   20-40\% centrality,
 (Right)  50-80\% centrality .
Solid circles are raw signals. Dash dotted blue lines are harmonic
combinational backgrounds constructed using formula (1) and dash black lines are
backgrounds constructed using formula (4).}
\label{200GeV_dihadron}
\end{figure*}

Normally, the dihadron combinational background can be described by the formula as follows~\cite{Ko,GL}:

\begin{widetext}
\begin{equation}
\left \langle f(\Delta \phi ) \right \rangle_{e}=\left \langle
\frac{N^{trig}\cdot N^{assoc}}{2\pi } \right \rangle_{e}+\left \langle
\frac{N^{trig}\cdot N^{assoc}}{2\pi }\cdot 2\sum_{n=1}^{+\infty}\upsilon
_{n}^{trig}\cdot \upsilon _{n}^{assoc} \right \rangle_{e}cosn\Delta \phi
\label{eqn1}
\end{equation}
\end{widetext}
where suffix $e$ in $\left \langle ...\right \rangle_{e}$ stands for
event-averaged quantity, while quantities without suffixes are for each event.

This is the first method, which is adopted in this study~\cite{GL}. By applying the method,  the initial geometry
anisotropy $\upsilon_{n}^{r}$ is obtained with respect to the event plane
extracted from the
stage before the parton cascade. This will help to exclude as much non-flow
contribution as possible. The event plane angle
of each harmonic order is calculated by using the following formula:
\begin{equation}
\Psi_{n}^{r}=\frac{1}{n}\left [ arctan\frac{\left \langle r^{n}sin(n\phi)
    \right \rangle}{\left \langle r^{n}cos(n\phi ) \right \rangle} +\pi \right
    ],
    \end{equation}
where $r$ and $\phi$ are the polar coordinates of partons in the initial coordinate space.
Then $\upsilon_{n}^{assoc}$ and $\upsilon_{n}^{trig}$ are calculated by using the
formula:
\begin{equation}
\upsilon _{n}^{r}=\left \langle cos\left [ n(\phi -\Psi _{n}^{r}) \right ]
\right \rangle,
\end{equation}
where $\phi$ is the azimuthal angle in the final state momentum
space. Here resolution correction for $\Psi _{n}^{r}$ is not necessary, since the
event plane resolution from parton configuration space is considered to be nearly 100\%.
As discussed earlier, this procedure effectively eliminates much nonflow contributions to
harmonic flows~\cite{Han}. From Ref.~\cite{GL}, we have seen that even in
the most central collision (b = 0 fm), it is better to include
higher order flow (up to fifth order) in background construction.
Therefore, we have included up to fifth order flow contribution in
background reconstruction.

However, here is a crucial point, which can not be neglected at the present stage.  If one multiplies
$\upsilon_{n}^{assoc}$ and $\upsilon_{n}^{trig}$ before
being event averaged, two-particle $\upsilon_{n}$ correlation can lead to a contribution
to the background, i.e. whether the factorization ($\left \langle \upsilon_{n}^{assoc}  \upsilon_{n}^{trig} \right \rangle_{e}$= $\left \langle \upsilon_{n}^{assoc} \right \rangle_{e} \left \langle\upsilon_{n}^{trig} \right \rangle_{e}$) is held or not.  In order to recover from the problem, the other modified form of the
formula is also put forward in the literature \cite{STAR-dihadron-1,STAR-dihadron-2}.  In the modified formula,
        systematic errors on $\upsilon_{n}$ are applied using background
        (Eq.\ref{eqn1}) as the upper bound. It can be written as:

\begin{widetext}
\begin{equation}
\left \langle f(\Delta \phi ) \right \rangle_{e}=\left \langle
\frac{N^{trig}\cdot N^{assoc}}{2\pi } \right \rangle_{e}+\left \langle
\frac{N^{trig}\cdot N^{assoc}}{2\pi } \right
\rangle_{e}2\sum_{n=1}^{+\infty}\left \langle \upsilon _{n}^{trig}\right
\rangle_{e}\left \langle \upsilon _{n}^{assoc} \right \rangle_{e}cosn\Delta\phi .
\label{eqn4}
\end{equation}
\end{widetext}

The difference between Eqns (\ref{eqn1}) and (\ref{eqn4}) contains two kinds of contributions: 1. Flow
fluctuation and its correlation from initial geometry asymmetry;
2. nonflow and its correlation. The first kind of contribution
should be included in background, while the second kind should be
excluded. However, neither background (Eq.1) and nor background (Eq.4) can
meet the demand. The background (Eq.1) is found to overestimate background by
including nonflow contribution, while background (Eq.4) underestimates
the background by throwing away the background from  flow
fluctuation and its correlation.  It is worth mentioning that
there have been some efforts to solve the crucial problem on
decomposing flow,
 flow fluctuation, and nonflow~\cite{purduenonflow}.

Although an ideal background is hard to be obtained, however, one can use the two
formulas from Eqns. (1) and (4)  as the upper and lower limits of the background to
get a reasonable range of jet-medium contribution. Following the same method, background is reconstructed using Eqns. (1) and (4)
as the upper and lower limits for dihadron background.
Correspondingly, they are marked as ``background (Eq.1)" and ``background (Eq.4)" in
the figures. The detailed values of parameters $\left
\langle\upsilon_{1}^{trig}\right \rangle$, $\left
\langle\upsilon_{1}^{assoc}\right \rangle$ and $\left \langle\upsilon_{1}^{trig} \upsilon_{1}^{assoc} \right \rangle$ used in these two methods of background
extractions are listed in appendix A.

From the Eqns. (1) and (4), we have observed that the normalization
factor of the background should be $\left \langle
\frac{N^{trig}\cdot N^{assoc}}{2\pi } \right \rangle_{e}$
theoretically. Another method is to use ZYAM scheme (A Zero Yield At Minimum)
for adjusting this factor to best match the signal
with experimental findings. However, we have checked the results from both way and found that  the difference between
the theoretical normalization factor and  the ZYAM adjustment is
less than 3\%.

\begin{figure*}[hbtp]
\includegraphics[width=6.8cm]{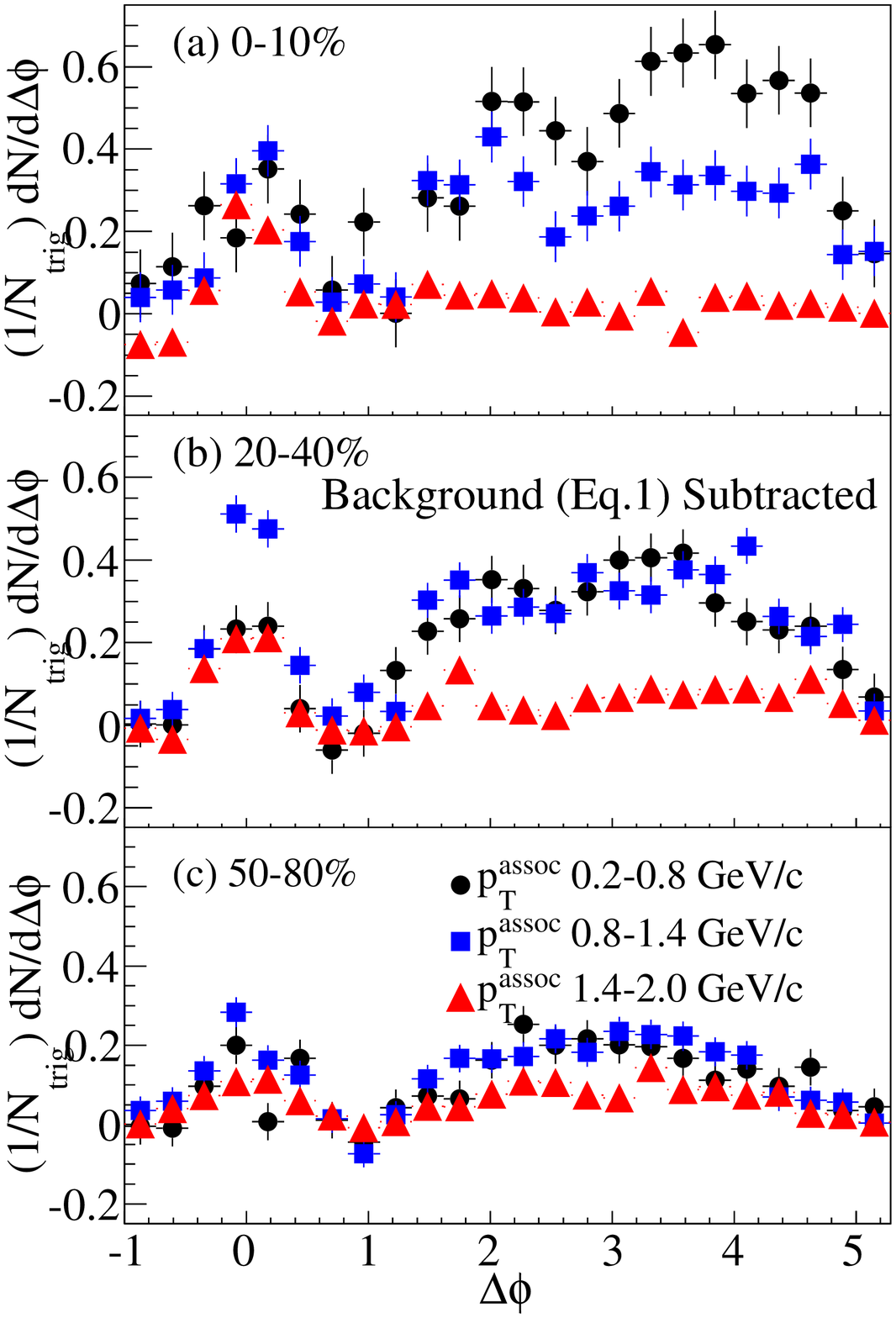}
\includegraphics[width=6.8cm]{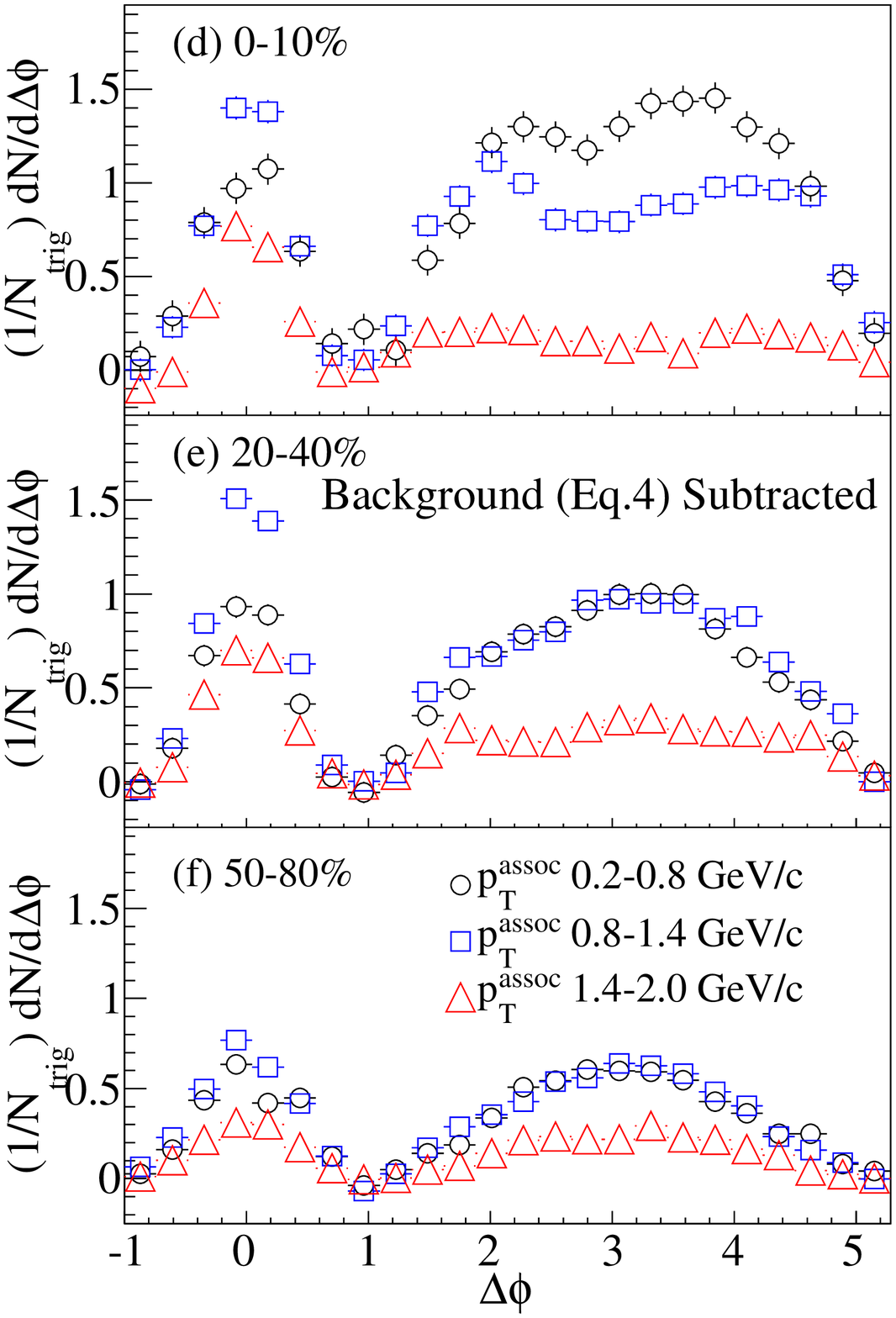}
\caption{(Color online) Background subtracted signals in three $p_{T}^{assoc}$ bins for three centralities.
Panels (a), (b), (c) stand for the signals with background (Eq.1) subtracted;
panels (d), (e), (f) stands for the signals after background (Eq.4) subtracted. See texts for details.
}
\label{200GeV_dihadron_real_bg1}
\end{figure*}

\begin{figure*}[hbtp]
\includegraphics[width=6.8cm]{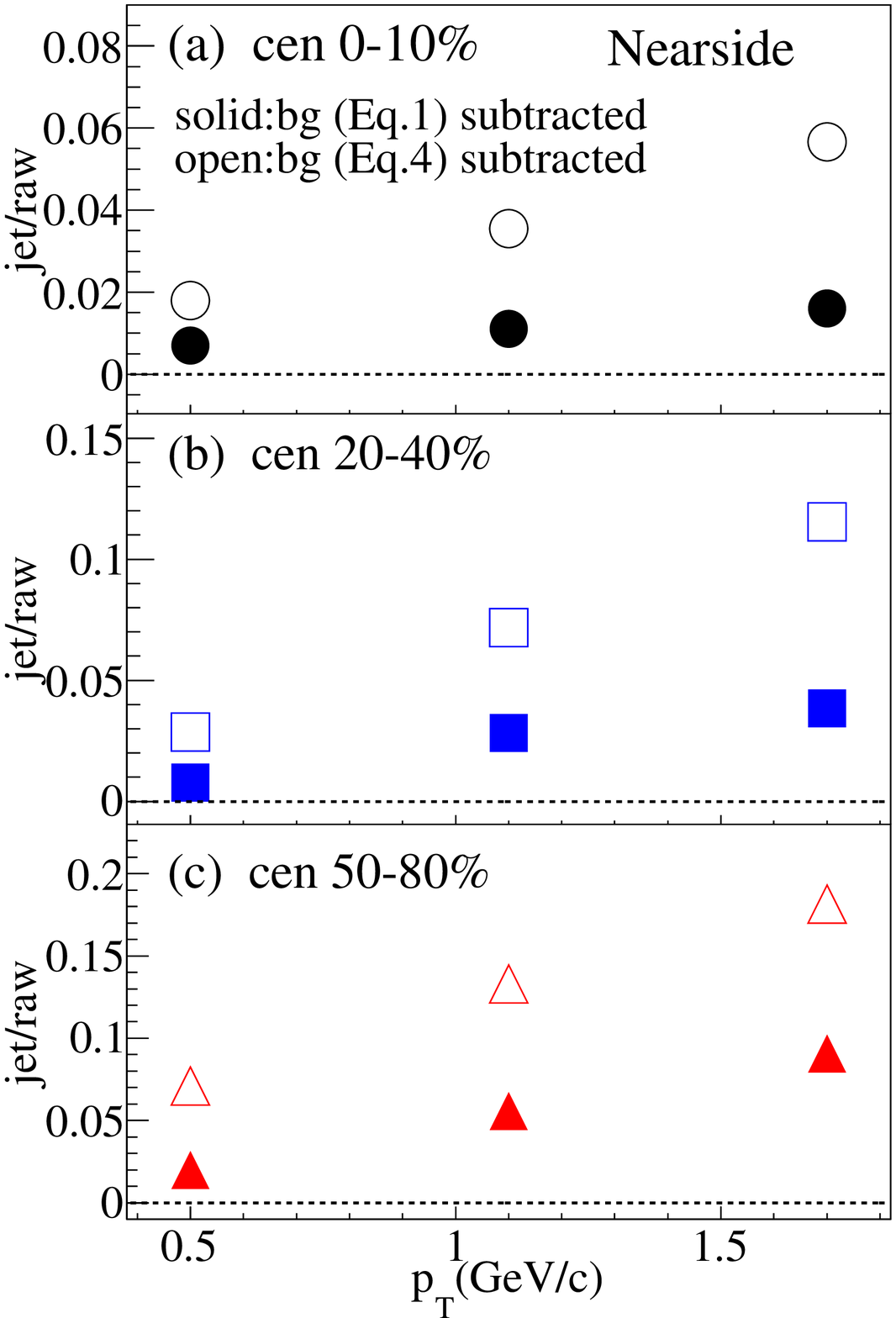}
\includegraphics[width=6.8cm]{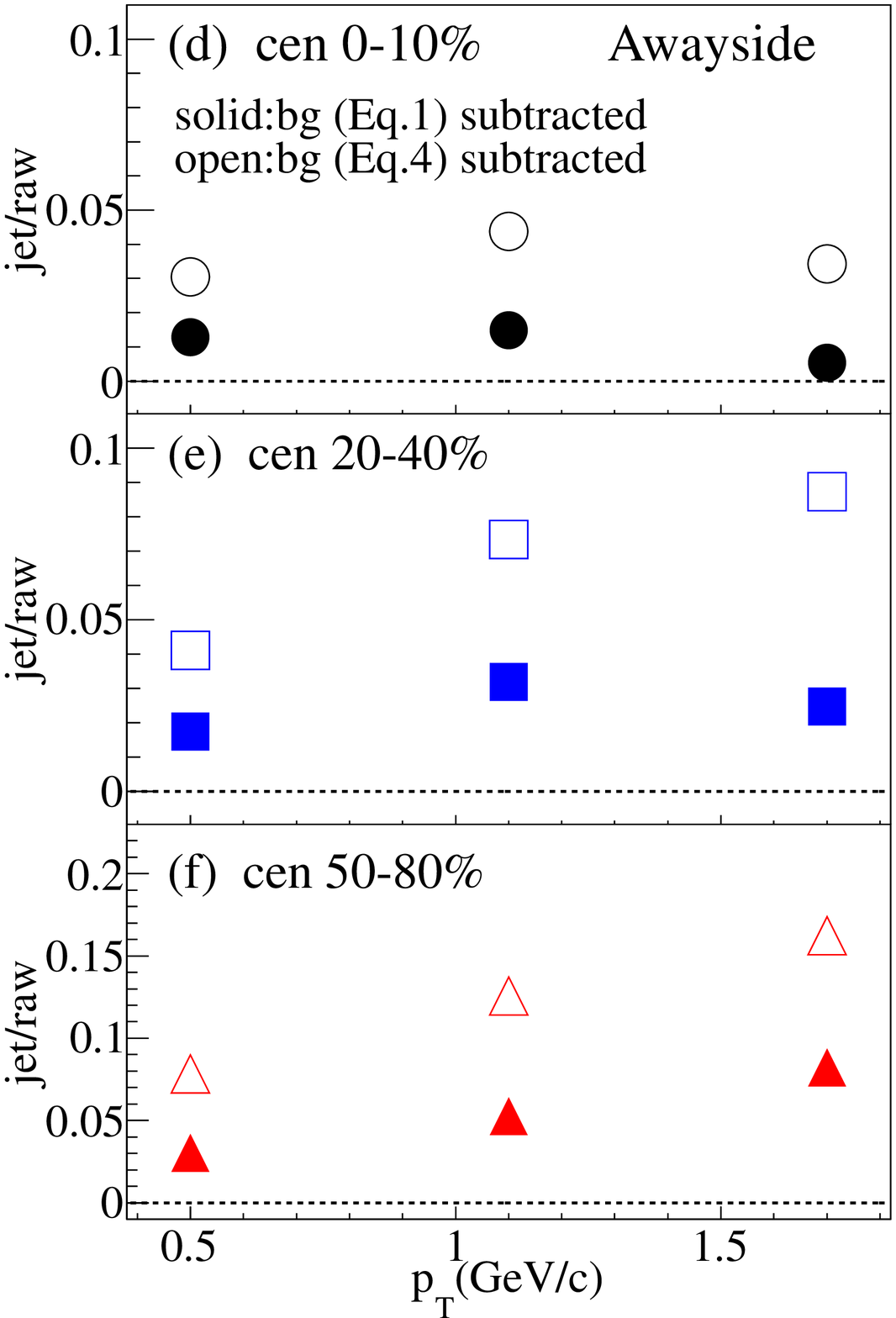}
\caption{(Color online) Jet relative contributions as a function of
    $p_{T}^{assoc}$ in three centrality bins for near-side (panels:(a),(b),(c)) and away-side (panels:(d),(e),(f)).
}
\label{200GeV_dihadron_realratio_near}
\end{figure*}

\section{Results and Discussions}

\subsection{Two Particle Azimuthal Correlation in Three Centrality Bins}

Fig. 1 shows the dihadron azimuthal correlations (both raw and combined
harmonic background) with different centrality bins from left to right panels. Each panel from top to bottom is
displaying the results with different $p_{T}^{assoc}$ bins.
For the better understanding, two backgrounds have been drawn simultaneously.

Firstly, an obvious difference in shape of background in different
centralities is observed. The central 0-10\% events have flatter or even double-peak shape
background, while 20-40\% and 50-80\% ones have a single-Gaussian shape
background. This is due to the different propagation properties of harmonic flows for different centralities.
The existence of the hot dense matter (QGP) propagates the initial geometry
irregularities to a larger extent. In Ref. \cite{Han}, $p_T$ dependencies of $v_2$ and
$v_3$ show a stronger dependence especially in large $p_T$ range,
which makes $\upsilon_{3}$ to increase more rapidly. Furthermore, $\upsilon_{3}$ contribution is
strongly affected with the change in the geometry from central to peripheral one. This leads to the
change of combinational background shape to  flatter or  double peak
in the $p_{T}$ range from 1.4 to 2.0 GeV/c.  For the two backgrounds, they tend to have similar shapes and just differ a little bit in magnitudes.

Secondly, we have found a seemingly changing trend of the signal using either
background. Therefore, we need to subtract the background and further study the jet-medium contribution. The
corresponding plots are shown in Fig. 2. The panels (a) to (c) ((d) to (f)) represent the results with background Eq.1 (Eq.4) subtracted.
From the global comparison of these two panel sets, an obvious change in signal
shape has been observed. The signals in more central events (0-10\%) or higher $p_{T}^{assoc}$
range (eg. 1.4-2.0 GeV/c) tend to have flatter shape signal, while peripheral
50-80\% events almost show a single-peak shape signal. This difference is consistent with the medium modification picture. The
existence of QGP strongly modifies jets, which makes the correlation shape
flatter (or more broaden) in the most central collisions. In addition, it also suppresses higher
$p_{T}$ particles. On the other hand, if jets interactions with surrounding medium particles are less (peripheral collisions),
the correlation shape will tend to be single Gaussian.
Moreover, the previous results indicate that there is also some contributions from hot spots, when the jet production is switched off~\cite{GL}.

For the further analysis, a quantity named ``jet relative contribution" is used to
represent the contribution of jet-medium correlation in total dihadron
correlation function. It is defined as the ratio of jet-medium correlation function yield to  the raw
dihadron correlation yield (including flow background). These jet relative contributions
in different centrality and $p_{T}^{assoc}$ bins are calculated and results are shown
in Fig. 3.
The panels (a), (b), (c) in Fig. 3 are for near-side and panels (d), (e), (f) are for
away-side jet relative contribution. The jet relative contributions using two different
backgrounds are drawn together for  providing an upper and lower
limit. From panel (d), one can see that the away-side jet relative
contribution in central 0-10\% events drops in $p_{T}^{assoc}$
range from 1.4 to 2.0 GeV/c, which is different from the case of near-side contribution
(panel (a)). This is consistent with the high $p_{T}$ suppression
picture in QGP. The quantitative values are provided in Table I
and II. In conclusion, the away-side jet relative contribution is
less than 5\% in central 0-10\% events.

\begin{table}
\caption{\label{tab:example}Near-side Jet-medium Contribution}
\begin{ruledtabular}
\begin{tabular}{llll}
   & 0.2-0.8 GeV/c & 0.8-1.4 GeV/c & 1.4-2.0 GeV/c\\
    0-10\% & 0.7\%-1.8\% & 1.1\%-3.5\% & 1.6\%-5.7\%\\
    20-40\% & 0.8\%-2.9\% & 2.8\%-7.2\% & 3.8\%-11.5\%\\
    50-80\% & 2.0\%-7.1\% & 5.2\%-13.3\% & 9.1\%-18.1\%\\
\end{tabular}
\end{ruledtabular}
\end{table}

\begin{table}
\caption{\label{tab:example2}Away-side Jet-medium Contribution}
\begin{ruledtabular}
\begin{tabular}{llll}
   & 0.2-0.8 GeV/c & 0.8-1.4 GeV/c & 1.4-2.0 GeV/c\\
    0-10\% & 1.3\%-3.1\% & 1.5\%-4.4\% & 0.5\%-3.4\%\\
    20-40\% & 1.7\%-4.1\% & 3.2\%-7.3\% & 2.5\%-8.7\%\\
    50-80\% & 3.0\%-7.8\% & 5.2\%-12.6\% & 8.2\%-16.2\%\\
\end{tabular}
\end{ruledtabular}
\end{table}

\subsection{$p_{T}^{assoc}$ dependence of jet-medium contribution in central 0-10\% collisions}

\begin{figure*}[hbtp]
\includegraphics[width=5.8cm]{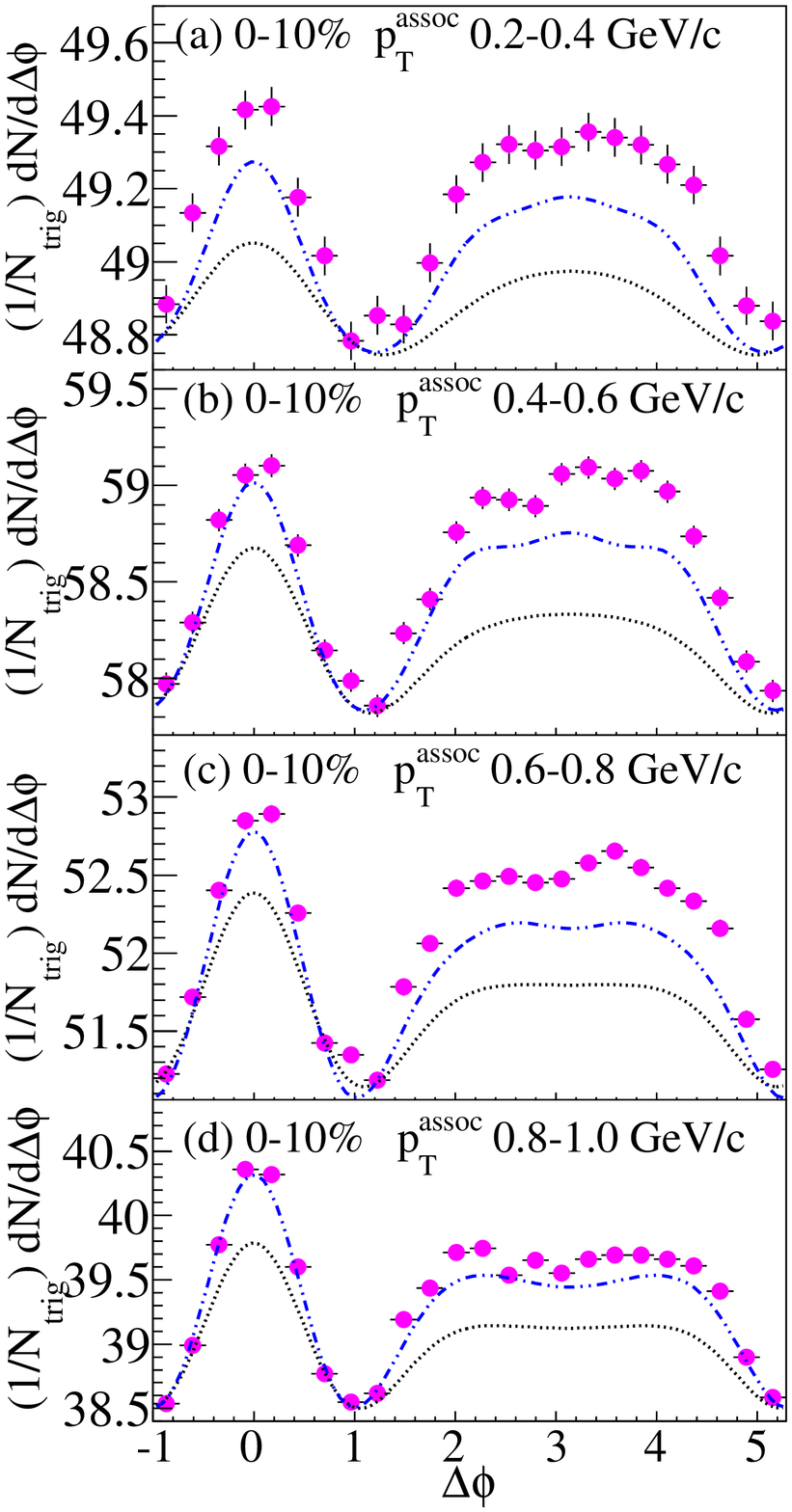}
\includegraphics[width=5.8cm]{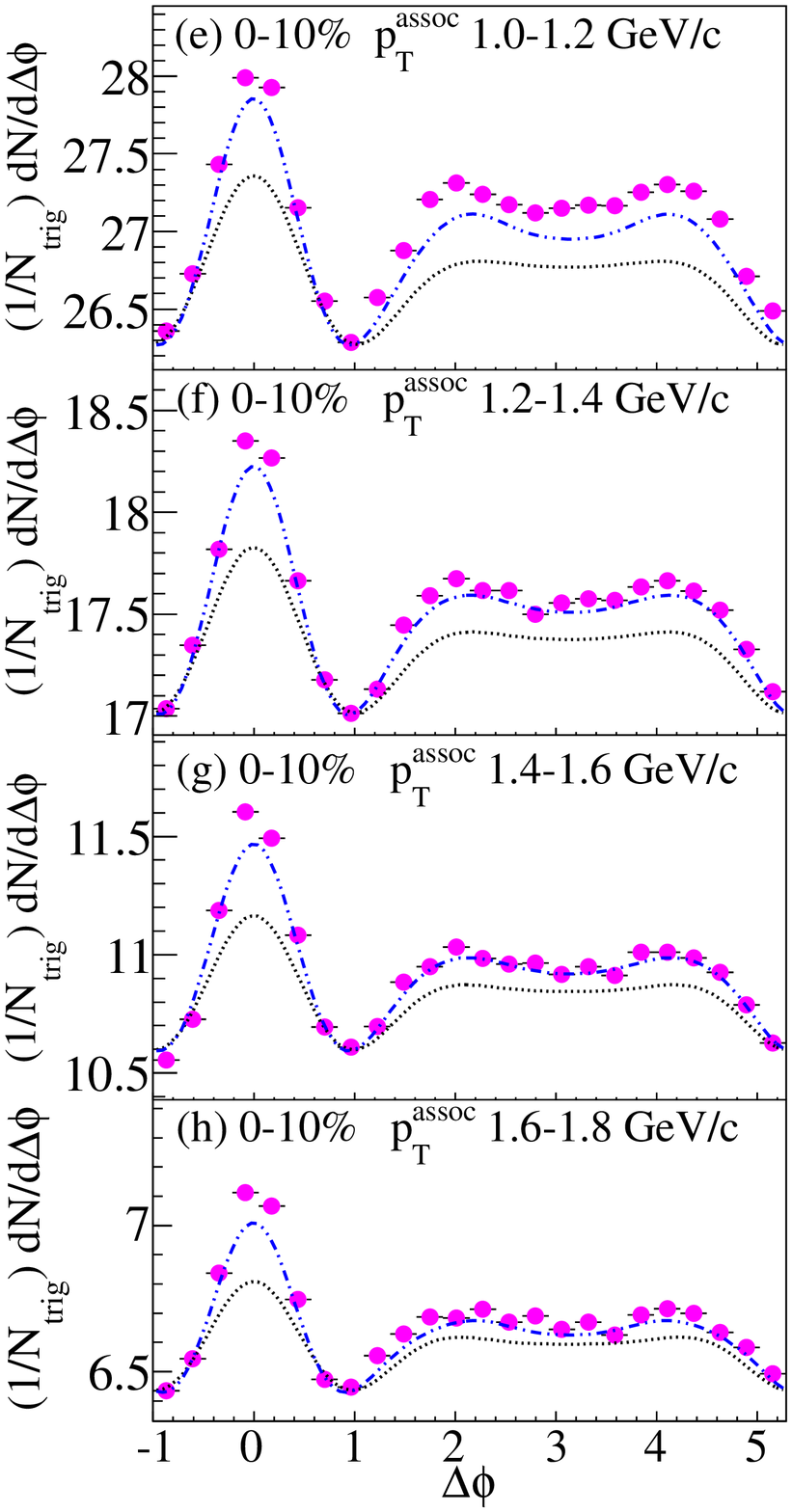}
\includegraphics[width=5.8cm]{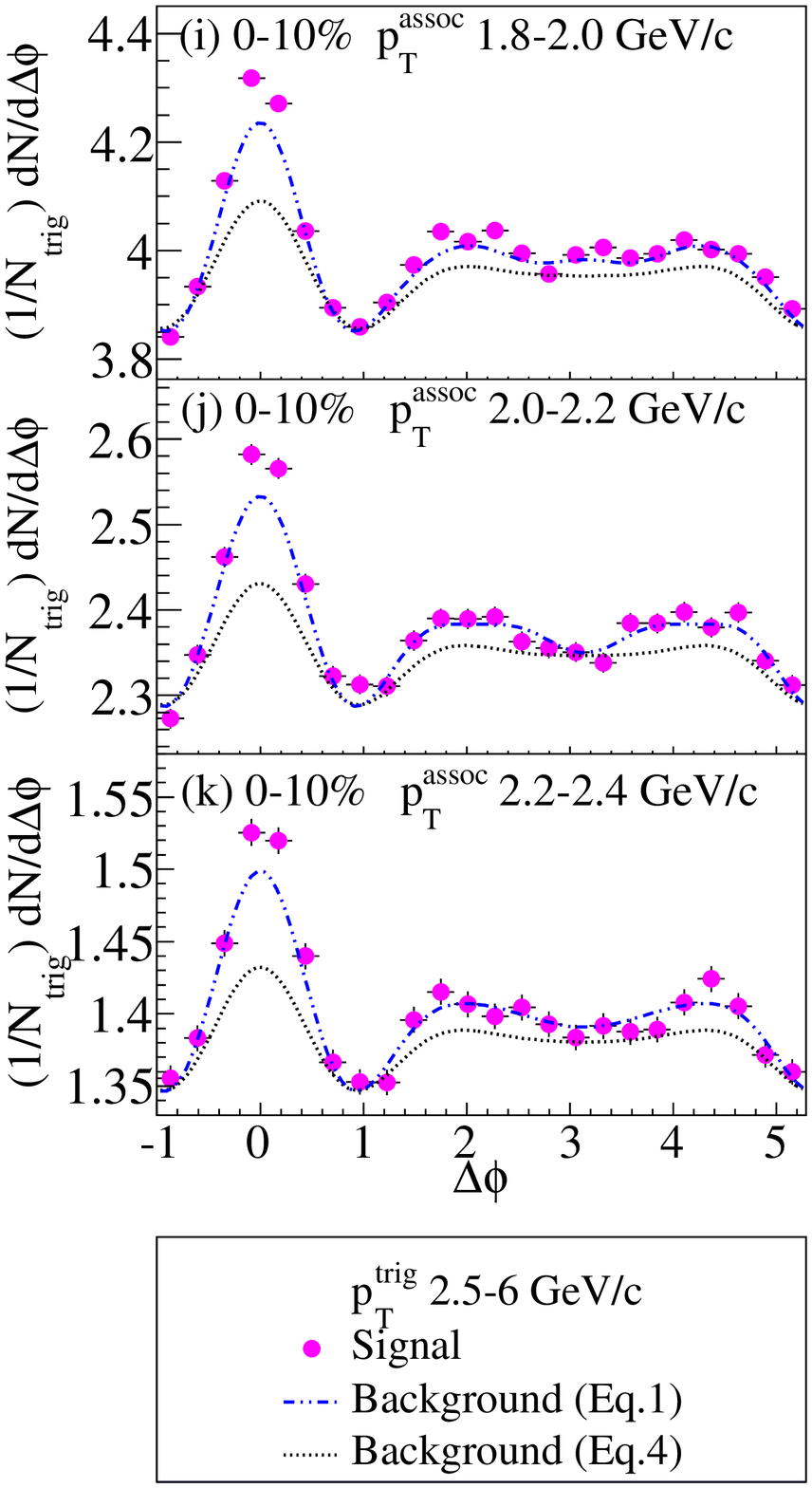}
\caption{(Color online) Dihadron correlation functions for different $p_{T}^{assoc}$ bins in 200 GeV/c
Au+Au collisions at 0-10\% centrality. From upper-left to bottom-right, the  $p_{T}^{assoc}$ are separated into
11 bins with 0.2 GeV/c bin width. Solid circles are raw signals, dash-dotted blue curves are the combinational
background (Eq.1) cases and dotted black curves are the combinational background (Eq.4) cases.}
\label{200GeV_dihadron}
\end{figure*}

\begin{figure*}[htbp]
\includegraphics[width=5.8cm]{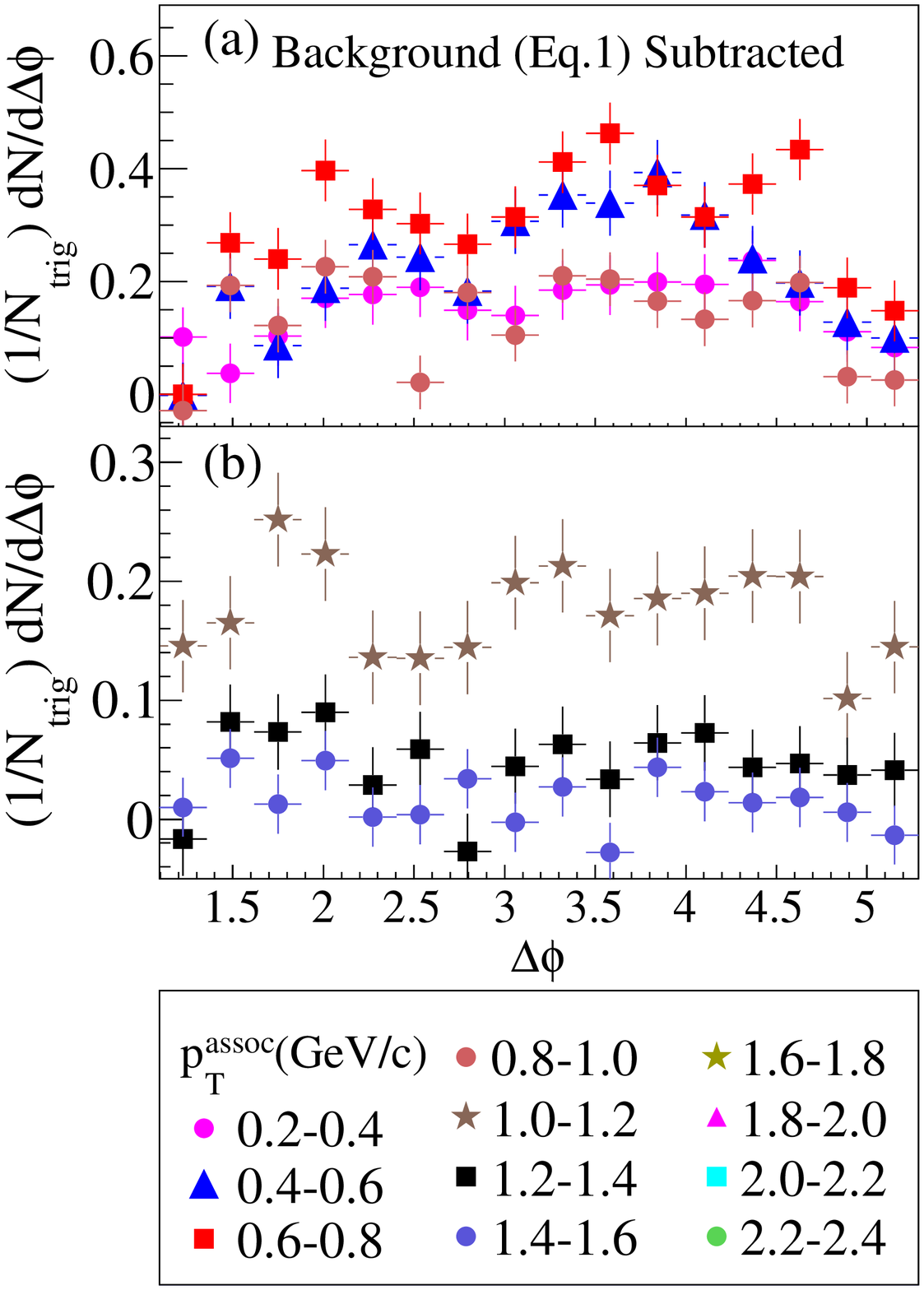}
\includegraphics[width=5.8cm]{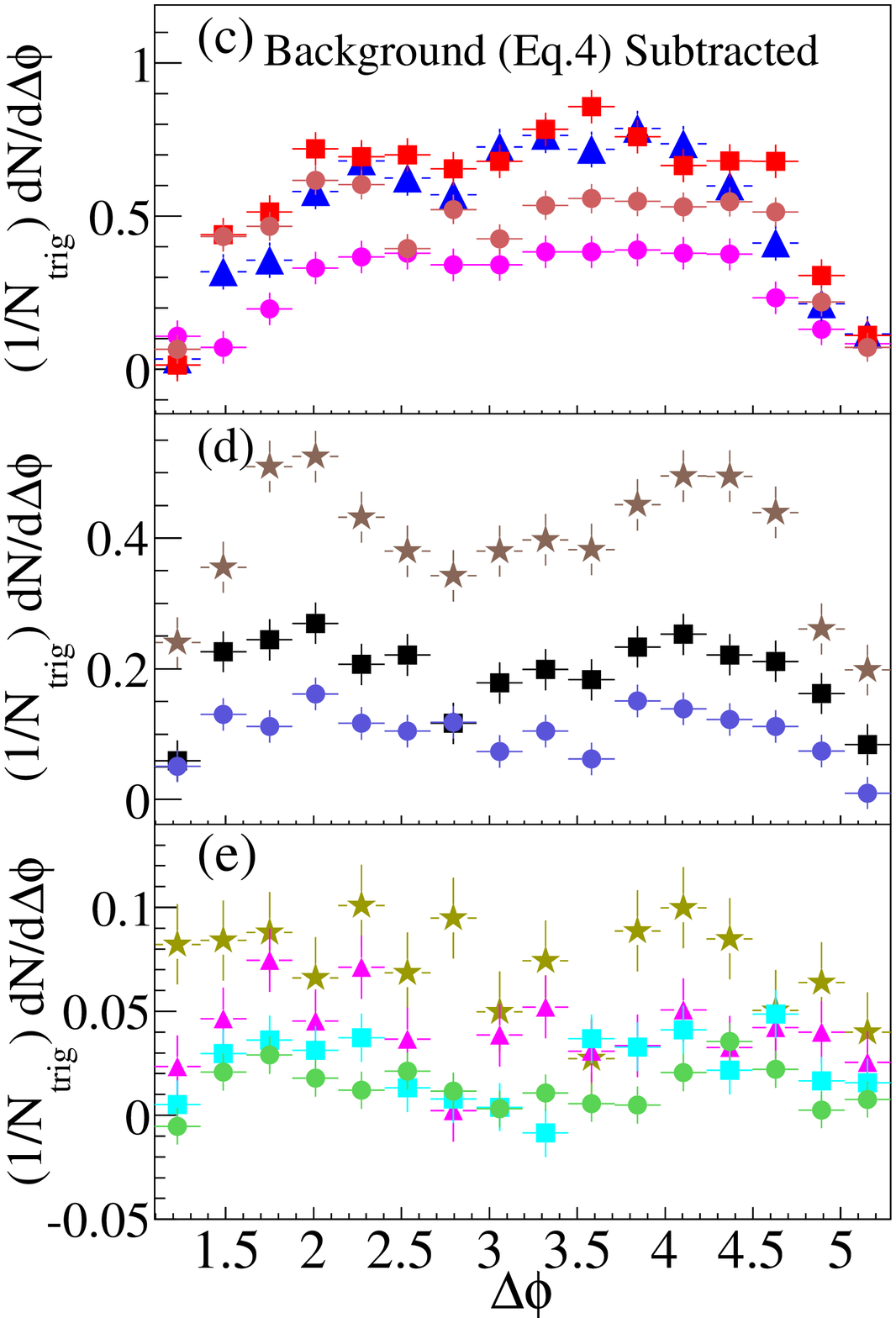}
\includegraphics[width=5.8cm]{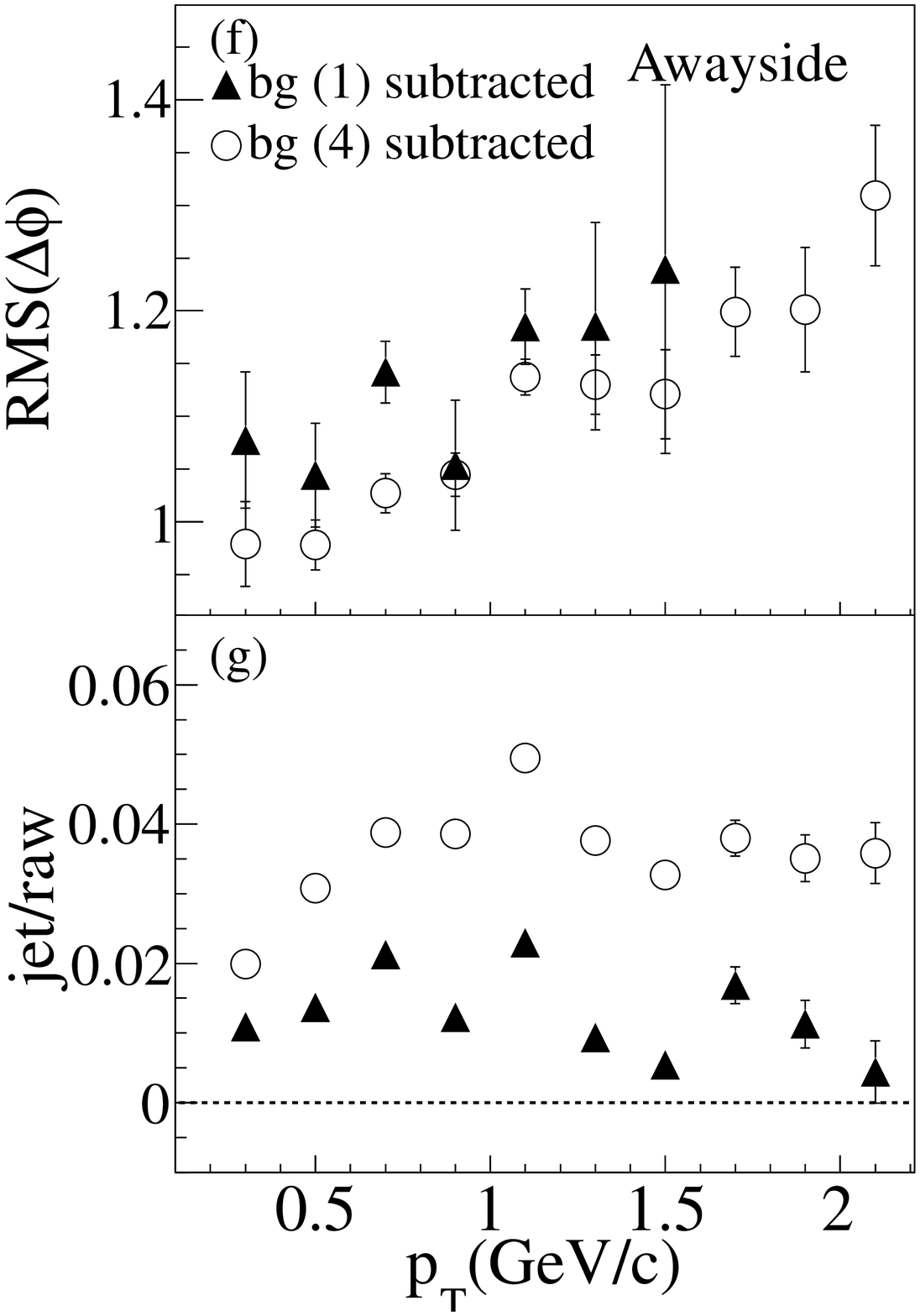}
\caption{(Color online) Panels (a), (b): background subtracted away-side signals for seven different
$p_{T}^{assoc}$ bins for background (Eq.1) case; Panels (c), (d), (e): away-side signals for eleven
different bins for background (Eq.4) case; Panels (f), (g): $p_{T}^{assoc}$ dependences
of RMS width and jet relative contribution.}
\label{200GeV_jet_real_9pT}
\end{figure*}

As we have already observed the modification to the correlation
function by jet-medium interactions in Fig. 2 and 3, now we would like to
focus on central 0-10\% events for more detailed
analysis. This is important because the modifications are the most obvious near the central geometry.
For this purpose, the whole $p_{T}^{assoc}$ range (from 0.2 GeV/c to 2.4 GeV/c) is divided into much
fine bins (with  a bin width of 0.2 GeV/c). The results are shown in Fig. 4, where total 11 $p_{T}^{assoc}$ bins are
taken into consideration for central events.
As is seen clearlt from Fig. 4 that the background (Eq.1) almost overlaps with raw signal in high $p_{T}$ range
(larger than 1.6 GeV/c), so we only pick out 7 $p_{T}$ bin results for further analysis.
For background (Eq.2) case, all the real signals are extracted.

The extracted values of background (Eq. 1) and background (Eq. 4) from Fig. 4 are displayed in the left and middle panels of Fig. 5.
The different colored and type of the symbols represent the values extracted from different
$p_{T}^{assoc}$ bins. For background (Eq.1) case (panels (a), (b)), we have found
that the signal shape becomes flatter and flatter towards the
high $p_{T}^{assoc}$. For background (Eq.4) case (panels (c) to (e)), a clear
evolution from a flat or seemingly single peak structure to a
double peak structure with $p_{T}^{assoc}$ is obtained. This is because of the
unsubtracted $\upsilon _{3}$ fluctuation and correlation. The RMS and jet relative contribution, which are extracted from
the left and middle panels, are shown in the right panels (f) and (g) of Fig. 5. The
$p_{T}^{assoc}$ dependence of RMS also shows the similar
evolutions of away-side signal shape for both the cases. The
reasonable range of jet relative contribution is shown in the
panel (g). These results suggest the
importance of jet-medium interactions, and may help elucidate the
mechanisms of jet energy loss in the QGP.

\section{Summary}
In summary, we have studied the dihadron azimuthal correlation functions  using a
multi-phase transport (AMPT) model in 200 GeV/c Au+Au collisions for different centralities of 0-10\%, 20-40\% and 50-80\%
and different $p_{T}^{assoc}$ bins. We have obtained the harmonic flows with less nonflow effect and constructed the combined harmonic
flow background using two formulas as reasonable upper and lower limits. Although the backgrounds calculated by two formulas differ in magnitude, the
physics information is quite similar for both cases.

The evolution of real signal shape
and away-side jet relative contribution with the increase of $p_{T}^{assoc}$
and centrality is consistent with the fact that the high $p_{T}$
particles are strongly modified by the hot dense medium and the hot dense medium formation is weaker in more peripheral collisions.

The jet contribution percentage in the raw dihadron correlation function is
relatively small. For the most central events (0-10\%), it is less than
5\%. For a reliable extraction of the jet-correlation yields, a precise
understanding of the harmonic flow background to within a few percent
is required. However, the jet-correlation shape is robust against a
large variation in the background subtraction. We have carried out a
comprehensive study of jet-medium contribution as a function of
$p_{T}^{assoc}$ in fine bins. We observe an evolution of the correlation
signal with increasing $p_{T}^{assoc}$, from a single Gaussian to a flat or
even double-peaked shape. The away-side correlation function RMS
increases with $p_{T}^{assoc}$, even for our conservative choices of the two
significantly different flow backgrounds. These results suggest the
importance of jet-medium interactions, and may help elucidate the
mechanisms of jet energy loss in the QGP.

This work was supported in part by the National Natural Science Foundation of
China under contract Nos. 11035009, 11220101005, 10979074,  11105207, 11175232, U1232206
and the Knowledge Innovation Project of the Chinese Academy of Sciences under Grant No. KJCX2-EW-N01.


\appendix
\section{List of values of flow parameters}
\begin{table*}
\caption{\label{tab:example}$\left \langle\upsilon_{1}^{trig}\right \rangle$  $\left \langle\upsilon_{1}^{assoc}\right \rangle$  $\left \langle\upsilon_{1}^{trig} \upsilon_{1}^{assoc} \right \rangle$}

\begin{ruledtabular}
\begin{tabular}{llllllllll}
   &  & 0-10\% &  &   & 20-40\% &  &  & 50-80\% &  \\
   $p_{T}^{assoc}$ range & $\left \langle\upsilon_{1}^{trig}\right \rangle$ & $\left \langle\upsilon_{1}^{assoc}\right \rangle$ & $\left \langle\upsilon_{1}^{trig} \upsilon_{1}^{assoc} \right \rangle$
   & $\left \langle\upsilon_{1}^{trig}\right \rangle$ & $\left \langle\upsilon_{1}^{assoc}\right \rangle$ & $\left \langle\upsilon_{1}^{trig} \upsilon_{1}^{assoc} \right \rangle$
   & $\left \langle\upsilon_{1}^{trig}\right \rangle$ & $\left \langle\upsilon_{1}^{assoc}\right \rangle$ & $\left \langle\upsilon_{1}^{trig} \upsilon_{1}^{assoc} \right \rangle$
\\
0.2-0.4 GeV/c &-0.025871 &0.003891 &-0.000697 &-0.002960 &0.001342 &-0.000942
&-0.005993 &0.000993 &-0.001940\\
        0.4-0.6 GeV/c &-0.025871 &0.004727 &-0.001506 &-0.002960 &0.001394
        &-0.001624 &-0.005993 &0.000642 &-0.002396\\
            0.6-0.8 GeV/c &-0.025871 &0.003769 &-0.001587 &-0.002960 &0.001312
            &-0.001456 &-0.005993 &-0.000149 &-0.002416\\
                0.8-1.0 GeV/c &-0.025871 &0.001927 &-0.001085 &-0.002960 &0.000702
                &-0.001577 &-0.005993 &-0.000068 &-0.002329\\
                    1.0-1.2 GeV/c &-0.025871 &-0.000342 &-0.000595 &-0.002960 &0.000570
                    &-0.001468 &-0.005993 &0.000488 &-0.003082\\
                        1.2-1.4 GeV/c &-0.025871 &-0.001999 &-0.000035 &-0.002960
                        &-0.000051 &-0.000855 &-0.005993 &-0.001384 &-0.003353\\
                            1.4-1.6 GeV/c &-0.025871 &-0.004586 &0.000807 &-0.002960
                            &-0.000390 &-0.000036 &-0.005993 &-0.003510 &-0.003151\\
                                1.6-1.8 GeV/c &-0.025871 &-0.007725 &0.001316 &-0.002960
                                &-0.001960 &0.000864 &-0.005993 &0.002034 &-0.003610\\
                                    1.8-2.0 GeV/c &-0.025871 &-0.010687 &0.002172 &-0.002960
                                    &-0.003233 &0.001257 &-0.005993 &0.000679 &-0.003284\\
                                        2.0-2.2 GeV/c &-0.025871 &-0.012145 &0.003433 &-0.002960
                                        &-0.002566 &0.001859 &-0.005993 &-0.002832 &-0.003654\\
                                            2.2-2.4 GeV/c &-0.025871 &-0.016067 &0.003415 &-0.002960
                                            &-0.003811 &0.002671 &-0.005993 &-0.005312 &-0.004666\\

\end{tabular}

\end{ruledtabular}
\end{table*}

\begin{table*}
\caption{\label{tab:example}$\left \langle\upsilon_{2}^{trig}\right \rangle$  $\left \langle\upsilon_{2}^{assoc}\right \rangle$  $\left \langle\upsilon_{2}^{trig} \upsilon_{2}^{assoc} \right \rangle$}

\begin{ruledtabular}
\begin{tabular}{llllllllll}
   &  & 0-10\% &  &   & 20-40\% &  &  & 50-80\% &  \\
   $p_{T}^{assoc}$ range & $\left \langle\upsilon_{2}^{trig}\right \rangle$ & $\left \langle\upsilon_{2}^{assoc}\right \rangle$ & $\left \langle\upsilon_{2}^{trig} \upsilon_{2}^{assoc} \right \rangle$
   & $\left \langle\upsilon_{2}^{trig}\right \rangle$ & $\left \langle\upsilon_{2}^{assoc}\right \rangle$ & $\left \langle\upsilon_{2}^{trig} \upsilon_{2}^{assoc} \right \rangle$
   & $\left \langle\upsilon_{2}^{trig}\right \rangle$ & $\left \langle\upsilon_{2}^{assoc}\right \rangle$ & $\left \langle\upsilon_{2}^{trig} \upsilon_{2}^{assoc} \right \rangle$
\\
0.2-0.4 GeV/c &0.073775 &0.016994 &0.001957 &0.188455 &0.041964 &0.009125
&0.164376 &0.052629 &0.012003\\
        0.4-0.6 GeV/c &0.073775 &0.031763 &0.003030 &0.188455 &0.070599 &0.015442
        &0.164376 &0.081647 &0.018957\\
            0.6-0.8 GeV/c &0.073775 &0.044547 &0.004151 &0.188455 &0.095192
            &0.021016 &0.164376 &0.103296 &0.024413\\
                0.8-1.0 GeV/c &0.073775 &0.053954 &0.004852 &0.188455 &0.114463
                &0.025449 &0.164376 &0.120846 &0.028254\\
                    1.0-1.2 GeV/c &0.073775 &0.060495 &0.005207 &0.188455 &0.129821
                    &0.029124 &0.164376 &0.133234 &0.032168\\
                        1.2-1.4 GeV/c &0.073775 &0.064775 &0.006021 &0.188455 &0.141029
                        &0.031751 &0.164376 &0.143701 &0.035081\\
                            1.4-1.6 GeV/c &0.073775 &0.067712 &0.006900 &0.188455 &0.149935
                            &0.034043 &0.164376 &0.151426 &0.038436\\
                                1.6-1.8 GeV/c &0.073775 &0.068329 &0.007220 &0.188455
                                &0.156176 &0.035729 &0.164376 &0.152973 &0.038768\\
                                    1.8-2.0 GeV/c &0.073775 &0.068417 &0.007574 &0.188455
                                    &0.160479 &0.037796 &0.164376 &0.155515 &0.040527\\
                                        2.0-2.2 GeV/c &0.073775 &0.066030 &0.007593 &0.188455
                                        &0.163173 &0.038348 &0.164376 &0.159956 &0.043669\\
                                            2.2-2.4 GeV/c &0.073775 &0.064178 &0.008366 &0.188455
                                            &0.163583 &0.038770 &0.164376 &0.160305 &0.043453\\
\end{tabular}

\end{ruledtabular}
\end{table*}

\begin{table*}
\caption{\label{tab:example}$\left \langle\upsilon_{3}^{trig}\right \rangle$  $\left \langle\upsilon_{3}^{assoc}\right \rangle$  $\left \langle\upsilon_{3}^{trig} \upsilon_{3}^{assoc} \right \rangle$}
\begin{ruledtabular}
\begin{tabular}{llllllllll}
   &  & 0-10\% &  &   & 20-40\% &  &  & 50-80\% &  \\
   $p_{T}^{assoc}$ range & $\left \langle\upsilon_{3}^{trig}\right \rangle$ & $\left \langle\upsilon_{3}^{assoc}\right \rangle$ & $\left \langle\upsilon_{3}^{trig} \upsilon_{3}^{assoc} \right \rangle$
   & $\left \langle\upsilon_{3}^{trig}\right \rangle$ & $\left \langle\upsilon_{3}^{assoc}\right \rangle$ & $\left \langle\upsilon_{3}^{trig} \upsilon_{3}^{assoc} \right \rangle$
   & $\left \langle\upsilon_{3}^{trig}\right \rangle$ & $\left \langle\upsilon_{3}^{assoc}\right \rangle$ & $\left \langle\upsilon_{3}^{trig} \upsilon_{3}^{assoc} \right \rangle$
\\
0.2-0.4 GeV/c &0.100672 &0.004909 &0.001118 &0.098937 &0.010104 &0.001703
&0.057041 &0.012910 &0.002508\\
        0.4-0.6 GeV/c &0.100672 &0.015469 &0.002593 &0.098937 &0.021127 &0.003401
        &0.057041 &0.019854 &0.003226\\
            0.6-0.8 GeV/c &0.100672 &0.028102 &0.004178 &0.098937 &0.032916
            &0.005453 &0.057041 &0.027854 &0.004895\\
                0.8-1.0 GeV/c &0.100672 &0.040093 &0.006086 &0.098937 &0.043845
                &0.007427 &0.057041 &0.033936 &0.006076\\
                    1.0-1.2 GeV/c &0.100672 &0.050931 &0.007859 &0.098937 &0.053737
                    &0.008962 &0.057041 &0.039496 &0.006865\\
                        1.2-1.4 GeV/c &0.100672 &0.059100 &0.008760 &0.098937 &0.061137
                        &0.010424 &0.057041 &0.045423 &0.008008\\
                            1.4-1.6 GeV/c &0.100672 &0.066054 &0.009964 &0.098937 &0.068773
                            &0.011919 &0.057041 &0.048002 &0.008809\\
                                1.6-1.8 GeV/c &0.100672 &0.071490 &0.010496 &0.098937
                                &0.074461 &0.013307 &0.057041 &0.051733 &0.010559\\
                                    1.8-2.0 GeV/c &0.100672 &0.075416 &0.011367 &0.098937
                                    &0.077711 &0.013558 &0.057041 &0.054062 &0.010938\\
                                        2.0-2.2 GeV/c &0.100672 &0.076608 &0.012745 &0.098937
                                        &0.082178 &0.014395 &0.057041 &0.058495 &0.008238\\
                                            2.2-2.4 GeV/c &0.100672 &0.078378 &0.013210 &0.098937
                                            &0.085642 &0.014619 &0.057041 &0.061674 &0.014042\\
\end{tabular}

\end{ruledtabular}
\end{table*}

\begin{table*}
\caption{\label{tab:example}$\left \langle\upsilon_{4}^{trig}\right \rangle$  $\left \langle\upsilon_{4}^{assoc}\right \rangle$  $\left \langle\upsilon_{4}^{trig} \upsilon_{4}^{assoc} \right \rangle$}
\begin{ruledtabular}
\begin{tabular}{llllllllll}
   &  & 0-10\% &  &   & 20-40\% &  &  & 50-80\% &  \\
   $p_{T}^{assoc}$ range & $\left \langle\upsilon_{4}^{trig}\right \rangle$ & $\left \langle\upsilon_{4}^{assoc}\right \rangle$ & $\left \langle\upsilon_{4}^{trig} \upsilon_{4}^{assoc} \right \rangle$
   & $\left \langle\upsilon_{4}^{trig}\right \rangle$ & $\left \langle\upsilon_{4}^{assoc}\right \rangle$ & $\left \langle\upsilon_{4}^{trig} \upsilon_{4}^{assoc} \right \rangle$
   & $\left \langle\upsilon_{4}^{trig}\right \rangle$ & $\left \langle\upsilon_{4}^{assoc}\right \rangle$ & $\left \langle\upsilon_{4}^{trig} \upsilon_{4}^{assoc} \right \rangle$
\\
0.2-0.4 GeV/c &0.075574 &0.000948 &0.000221 &0.030574 &0.001625 &0.000085
&-0.003253 &-0.001360 &0.000677\\
        0.4-0.6 GeV/c &0.075574 &0.005678 &0.000723 &0.030574 &0.004099 &0.000683
        &-0.003253 &-0.002469 &0.000986\\
            0.6-0.8 GeV/c &0.075574 &0.012493 &0.001632 &0.030574 &0.007566
            &0.001628 &-0.003253 &-0.002682 &0.001783\\
                0.8-1.0 GeV/c &0.075574 &0.020305 &0.002501 &0.030574 &0.011671
                &0.002133 &-0.003253 &-0.003266 &0.002189\\
                    1.0-1.2 GeV/c &0.075574 &0.028170 &0.003525 &0.030574 &0.015414
                    &0.003059 &-0.003253 &-0.002346 &0.003384\\
                        1.2-1.4 GeV/c &0.075574 &0.035320 &0.004594 &0.030574 &0.018721
                        &0.004194 &-0.003253 &-0.002007 &0.002937\\
                            1.4-1.6 GeV/c &0.075574 &0.042858 &0.005170 &0.030574 &0.021235
                            &0.004439 &-0.003253 &-0.002812 &0.002656\\
                                1.6-1.8 GeV/c &0.075574 &0.048819 &0.005748 &0.030574
                                &0.025388 &0.005695 &-0.003253 &-0.004228 &0.001316\\
                                    1.8-2.0 GeV/c &0.075574 &0.053568 &0.006991 &0.030574
                                    &0.028373 &0.006568 &-0.003253 &-0.005650 &0.002296\\
                                        2.0-2.2 GeV/c &0.075574 &0.056891 &0.006922 &0.030574
                                        &0.029028 &0.006538 &-0.003253 &-0.002869 &0.005727\\
                                            2.2-2.4 GeV/c &0.075574 &0.058573 &0.007499 &0.030574
                                            &0.031020 &0.005983 &-0.003253 &-0.000363 &0.005555\\
\end{tabular}

\end{ruledtabular}
\end{table*}

\begin{table*}
\caption{\label{tab:example}$\left \langle\upsilon_{5}^{trig}\right \rangle$  $\left \langle\upsilon_{5}^{assoc}\right \rangle$  $\left \langle\upsilon_{5}^{trig} \upsilon_{5}^{assoc} \right \rangle$}
\begin{ruledtabular}
\begin{tabular}{llllllllll}
   &  & 0-10\% &  &   & 20-40\% &  &  & 50-80\% &  \\
   $p_{T}^{assoc}$ range & $\left \langle\upsilon_{5}^{trig}\right \rangle$ & $\left \langle\upsilon_{5}^{assoc}\right \rangle$ & $\left \langle\upsilon_{5}^{trig} \upsilon_{5}^{assoc} \right \rangle$
   & $\left \langle\upsilon_{5}^{trig}\right \rangle$ & $\left \langle\upsilon_{5}^{assoc}\right \rangle$ & $\left \langle\upsilon_{5}^{trig} \upsilon_{5}^{assoc} \right \rangle$
   & $\left \langle\upsilon_{5}^{trig}\right \rangle$ & $\left \langle\upsilon_{5}^{assoc}\right \rangle$ & $\left \langle\upsilon_{5}^{trig} \upsilon_{5}^{assoc} \right \rangle$
\\
0.2-0.4 GeV/c &0.035613 &0.000163 &0.000072 &-0.004876 &-0.000210 &0.000189
&-0.006631 &-0.001752 &-0.000043\\
        0.4-0.6 GeV/c &0.035613 &0.001217 &0.000029 &-0.004876 &-0.000709
        &0.000279 &-0.006631 &-0.003521 &-0.000147\\
            0.6-0.8 GeV/c &0.035613 &0.003841 &0.000401 &-0.004876 &-0.000459
            &0.000549 &-0.006631 &-0.004073 &0.000861\\
                0.8-1.0 GeV/c &0.035613 &0.006993 &0.000565 &-0.004876 &-0.000890
                &0.000854 &-0.006631 &-0.004225 &0.000594\\
                    1.0-1.2 GeV/c &0.035613 &0.010640 &0.001154 &-0.004876 &-0.000437
                    &0.000669 &-0.006631 &-0.005513 &-0.000439\\
                        1.2-1.4 GeV/c &0.035613 &0.014408 &0.001556 &-0.004876 &-0.000367
                        &0.000239 &-0.006631 &-0.005430 &0.001545\\
                            1.4-1.6 GeV/c &0.035613 &0.018030 &0.001783 &-0.004876 &0.000923
                            &0.000809 &-0.006631 &-0.006749 &0.001939\\
                                1.6-1.8 GeV/c &0.035613 &0.020953 &0.002676 &-0.004876
                                &0.002621 &0.001746 &-0.006631 &-0.007577 &0.002326\\
                                    1.8-2.0 GeV/c &0.035613 &0.023771 &0.002333 &-0.004876
                                    &0.001566 &0.001071 &-0.006631 &-0.005370 &0.001255\\
                                        2.0-2.2 GeV/c &0.035613 &0.027007 &0.003120 &-0.004876
                                        &0.001490 &0.002077 &-0.006631 &-0.011794 &0.001494\\
                                            2.2-2.4 GeV/c &0.035613 &0.030243 &0.002670 &-0.004876
                                            &0.003509 &0.002326 &-0.006631 &-0.011982 &0.001287\\
\end{tabular}

\end{ruledtabular}
\end{table*}

\begin{table*}
\caption{\label{tab:example}$\left \langle\upsilon_{6}^{trig}\right \rangle$  $\left \langle\upsilon_{6}^{assoc}\right \rangle$  $\left \langle\upsilon_{6}^{trig} \upsilon_{6}^{assoc} \right \rangle$}
\begin{ruledtabular}
\begin{tabular}{llllllllll}
   &  & 0-10\% &  &   & 20-40\% &  &  & 50-80\% &  \\
   $p_{T}^{assoc}$ range & $\left \langle\upsilon_{6}^{trig}\right \rangle$ & $\left \langle\upsilon_{6}^{assoc}\right \rangle$ & $\left \langle\upsilon_{6}^{trig} \upsilon_{6}^{assoc} \right \rangle$
   & $\left \langle\upsilon_{6}^{trig}\right \rangle$ & $\left \langle\upsilon_{6}^{assoc}\right \rangle$ & $\left \langle\upsilon_{6}^{trig} \upsilon_{6}^{assoc} \right \rangle$
   & $\left \langle\upsilon_{6}^{trig}\right \rangle$ & $\left \langle\upsilon_{6}^{assoc}\right \rangle$ & $\left \langle\upsilon_{6}^{trig} \upsilon_{6}^{assoc} \right \rangle$
\\
0.2-0.4 GeV/c &0.014519 &-0.000231 &0.000120 &-0.013741 &0.000010 &-0.000044
&-0.001509 &0.000120 &-0.000396\\
        0.4-0.6 GeV/c &0.014519 &0.000030 &0.000215 &-0.013741 &-0.000491
        &-0.000084 &-0.001509 &-0.001390 &-0.000110\\
            0.6-0.8 GeV/c &0.014519 &0.000156 &-0.000083 &-0.013741 &-0.001263
            &-0.000056 &-0.001509 &-0.001474 &-0.000271\\
                0.8-1.0 GeV/c &0.014519 &0.001058 &0.000036 &-0.013741 &-0.002195
                &0.000216 &-0.001509 &0.000550 &0.000498\\
                    1.0-1.2 GeV/c &0.014519 &0.002324 &0.000420 &-0.013741 &-0.002858
                    &0.000355 &-0.001509 &-0.000728 &0.000465\\
                        1.2-1.4 GeV/c &0.014519 &0.003721 &0.000349 &-0.013741 &-0.003420
                        &0.000420 &-0.001509 &-0.002572 &-0.000688\\
                            1.4-1.6 GeV/c &0.014519 &0.004106 &0.000447 &-0.013741
                            &-0.003858 &0.000966 &-0.001509 &-0.002692 &0.000093\\
                                1.6-1.8 GeV/c &0.014519 &0.006095 &0.000886 &-0.013741
                                &-0.005499 &0.000622 &-0.001509 &-0.002420 &0.002307\\
                                    1.8-2.0 GeV/c &0.014519 &0.007962 &0.001136 &-0.013741
                                    &-0.004095 &0.000763 &-0.001509 &-0.002748 &0.001520\\
                                        2.0-2.2 GeV/c &0.014519 &0.007611 &-0.000213 &-0.013741
                                        &-0.001545 &0.000804 &-0.001509 &0.002193 &-0.000750\\
                                            2.2-2.4 GeV/c &0.014519 &0.008142 &0.000370 &-0.013741
                                            &-0.001668 &0.000705 &-0.001509 &0.003055 &0.000971\\
\end{tabular}

\end{ruledtabular}
\end{table*}

\end{document}